\documentclass[twocolumn,showpacs,preprintnumbers,
amsmath,amssymb,aps,prd,nofootinbib,superscriptaddress,
eqsecnum]{revtex4}
\usepackage{graphicx}

\makeatletter
\renewcommand{\p@subsection}{}
\makeatother

\newcommand{\Slash}[1]{\ooalign{\hfil/\hfil\crcr$#1$}}
\def\la{\langle}
\def\ra{\rangle}

\begin{document}

\title{Conformal anomaly and the vector coupling in dense matter}

\author{Chihiro Sasaki}
\affiliation{%
Frankfurt Institute for Advanced Studies,
D-60438 Frankfurt am Main,
Germany
}

\author{Hyun Kyu Lee}
\author{Won-Gi Paeng}
\affiliation{%
Department of Physics, Hanyang University, Seoul 133-791, Korea
}

\author{Mannque Rho}
\affiliation{%
Institut de Physique Th\'eorique,
CEA Saclay, 91191 Gif-sur-Yvette c\'edex, France \& \\
Department of Physics, Hanyang University, Seoul 133-791, Korea
}

\date{\today}

\begin{abstract}
We construct an effective chiral Lagrangian for hadrons implemented
by the conformal invariance and discuss the properties of nuclear
matter at high density. The model is formulated based on two alternative
assignment, ``naive'' and mirror, of chirality to the nucleons.
It is shown that taking the dilaton limit, in which the mended symmetry
of Weinberg is manifest, the vector-meson Yukawa coupling becomes
suppressed and the symmetry energy becomes softer as one approaches the chiral phase transition. This leads to softer
equations of state (EoS) and could accommodate the EoS without any exotica
consistent with the recent measurement of a $1.97 \pm 0.04\,M_\odot$
neutron star.
\end{abstract}


\maketitle

\section{Introduction and Results}
\label{sec:int}
The state of cold dense matter in the vicinity of phase transition from baryonic matter to quark matter presumed to be present in the interior of compact stars is not understood at all. This is because there is no realistic model-independent tool to probe that regime. In this paper, inspired by  Weinberg's notion of ``mended symmetry"~\cite{weinberg,weinberg2}, we would like to explore the possibility that in baryonic matter at some high density, there emerge in the chiral limit a multiplet of massless particles consisting of Goldstone bosons as well as other massless particles to fill out a full representation of the chiral symmetry group of QCD. This issue is relevant not only for the phase structure of dense baryonic matter but also for understanding certain astrophysical properties of compact stars that are being observed. This has a potentially intriguing implication on the recently observed $1.97 \pm 0.04\,M_\odot$ neutron star~\cite{2solarmass} as will be explained in the concluding section.

Before entering into the details of our reasoning, we should underline our basic assumption. We will assume that as one approaches the chiral restoration point in density, local fields continue to be relevant degrees of freedom. There are at present neither strong theoretical arguments nor experimental indications for the validity of such an assumption. Should it turn out that the notion of local fields makes no sense at high density in the vicinity of the chiral phase transition, then what we present in what follows would have no value. If however the notion made sense, then the proposed scenario would have an important implication on what happens to the repulsive core, a long-standing mystery in nuclear physics and a crucial ingredient for the physics of compact stars. We will find that as one approaches the critical density, the repulsion should be strongly suppressed, a result which has not been previously uncovered.

In the broken symmetry sector, the symmetry of such multplets is not ``visible." Involving massless vector fields, what is at issue would then be (hidden) gauge symmetry manifesting explicitly at a possibly second order phase transition. In fact the hidden local symmetry (HLS) theory of Harada and Yamawaki~\cite{HLSloop} with the vector manifestation (VM) fixed point with the vector mesons joining the pions in the same multiplet is precisely of this class. One should note that the symmetry involved here is a flavor symmetry, which is of course not a fundamental symmetry contained in QCD. It should more appropriately be viewed as an ``emergent symmetry" -- analogously to the CP$^{N-1}$ model --, and as such can be extended, starting from 4D low-energy theorems, to an infinite tower of gauge fields leading to a deconstruction of the fifth dimension in 5D Yang-Mills theory~\cite{son}.\footnote{We note that such a Lagrangian arises also top-down from string theory~\cite{sakai-sugimoto}.} Whether or not and how the massless multplets can manifest themselves at a phase transition such as chiral restoration are totally unknown and constitute the main line of research in nuclear/hadron as well as astro-hadron physics.

The question we would like to raise here is: How to exploit the properties of hidden local symmetry in unraveling dense baryonic matter?\footnote{The same question was raised for high temperature, particularly, in connection with dilepton productions in heavy-ion collisions in \cite{BHHRS}.} For this purpose, we first note that there are two indispensable degrees of freedom that are missing in HLS Lagrangian, i.e., baryons and scalars. The HLS Lagrangian contains, apart from the pions, vector mesons but no scalars. In nuclear physics, as we know from Walecka model~\cite{walecka} that works fairly well for phenomena near nuclear matter density, together with the vector mesons $(\rho,\omega)$, a scalar meson  is indispensable, e.g., for binding. Now the scalar that figures in Walecka model cannot be the scalar of the linear sigma model, for if it were that scalar, nuclear matter would be unstable. In fact it has to be a chiral scalar. On the other hand, at high density, the relevant Lagrangian that has correct symmetry is the linear sigma model, and the scalar that is needed there is the fourth component of the chiral four-vector $(\pi_1,\pi_2,\pi_3,\sigma)$. Thus in order to probe highly dense matter, we have to figure out how the chiral scalar at low density transmutes to the fourth component of the four-vector. We should stress that this is a part of the long-standing scalar puzzle in low-energy hadron physics, which remains still highly controversial.

In this work, in the same spirit as what entered in the formulation of BR scaling~\cite{BR}, a chiral scalar will be introduced as a dilaton associated with broken conformal symmetry and responsible for the trace anomaly of QCD. Following \cite{miransky}, we write the trace anomaly -- which is proportional (in the chiral limit) to the gluon condensate $\la G_{\mu\nu}G^{\mu\nu}\ra$ -- in terms of ``soft" dilaton $\chi_s$ and ``hard" dilaton $\chi_h$. As suggested in \cite{LeeRho}, we will associate the soft dilaton with that component locked to the quark condensate $\la\bar{q}q\ra$. We assume that this is the component which ``melts" across the chiral phase transition, with the hard component remaining non-vanishing~\footnote{The ``melting" of the soft component is observed in dynamical lattice calculation in temperature~\cite{miller} but is an assumption in density.}.

As for baryons, the best way would be that they are generated as solitons in HLS theory. As stressed since a long time~\cite{CNDII}, baryons generated as skyrmions in the presence of vector mesons could most efficiently capture the strong-coupling physics needed for nuclear interactions both at nuclear matter density and at higher densities. Indeed this point is given a support by a recent calculation of finite nuclei in terms of BPS skyrmions obtained from an infinite-tower HLS Lagrangian where the higher tower is integrated out~\cite{sutcliffe}. It works much better than the standard skyrmion model without vector mesons in capturing the dynamics of few-body nuclear systems. It may be viewed as an additional support for the power of the HLS strategy advocated in \cite{CNDII}. Unfortunately a controlled systematic treatment of many-nucleon systems is mathematically involved and has not been worked out except for certain topologically robust properties~\cite{half} that will be mentioned below. We will therefore put nucleon fields by hand by coupling them in hidden gauge invariant way to the mesons $\pi$, $\rho$, $\omega$ and to the scalar dilaton.

In introducing baryonic degrees of freedom, there are two alternative ways of assigning chirality to the nucleons. One is the ``naive" assignment~\footnote{We put this terminology in a quotation mark since it is a misnomer, used merely to distinguish it from the alternative option.} and the other the mirror assignment. The ``naive" assignment is anchored on the standard chiral symmetry structure where the entire constituent quark or nucleon mass (in the chiral limit) is generated by spontaneous symmetry breaking. The nonlinear sigma model (and its gauge-equivalent HLS model) is the typical example of this type. The merit of this model is that it is consistent with the constituent quark model which enjoys the successful mass relation $m_B/m_\rho\simeq 3/2$ where $m_B$ is the average of the nucleon and $\Delta$ masses. The constituent quark model is supported by large $N_c$ considerations not only at low-energy scale but also at intermediate-energy scale~\cite{weinberg3}. Whether it is a viable model at shorter-scale as in high density is of course not known.

The alternative, mirror assignment~\cite{dk,mirror}, allows a chiral invariant mass term common to the parity-doublets that can remain non-zero at chiral restoration, which means that a part of the nucleon mass, say, $m_0$, must arise from a mechanism that is not associated with spontaneous chiral symmetry breaking. The origin of such a mass $m_0$ is not known, but \`a priori, there is no reason why it cannot be present. At present, analysis of various observables both in the vacuum such as pion-nucleon scattering etc. and in medium such as nuclear matter properties etc. based on linear and nonlinear sigma models with mirror symmetry~\cite{nemoto,analysis} cannot rule out an $m_0$ of a few hundred MeV. If $m_0$ is non-negligible, then the direct relation of masses between baryons and mesons enjoyed by the constituent quark model will no longer be obvious even if it still holds. Nonetheless there is a motivation for considering this scenario.

There is an unexpected indication from simulations of skyrmion matter on crystal lattice to introduce density that the meson and baryon masses behave differently in increasing density: the baryon mass appears to drop at a slower rate than the meson mass as density is increased and may not vanish at the chiral restoration point. This is an outcome of the model albeit at large $N_c$, not put in {\it ab initio}. Such a different in-medium behavior between mesons and baryons is found to have an important consequence on the nuclear tensor forces and hence on the EoS of baryonic matter at densities exceeding the nuclear matter density~\cite{LPR}. Given that the EoS involves shorter-length scale than that probed by vacuum and nuclear phenomenology, the mirror scenario combined with HLS may prove to be relevant for EoS at high density.

The objective of this paper is to explore the consequences of a dilaton-implemented HLS (dHLS for short) Lagrangian containing baryons both in the ``naive" and the mirror assignments at normal as well as high densities. The strategy we will use to drive the system from nuclear matter density to near chiral restoration density is the ``dilaton limit" proposed by Beane and van Kolck~\cite{vanKolck}.  Phenomenology of vacuum processes with this Lagrangian will be discussed elsewhere~\cite{PLRS}.

As a brief summary of the results, we note that the dHLS model at nuclear matter density in mean field in the ``naive" assignment scheme is equivalent to Walecka's mean-field model, with the scalar figuring as a chiral scalar. We expect that the same should hold in the mirror assignment if the model applied at normal nuclear matter density. As the dilaton limit is taken, a linear sigma model in both assignments emerges from the highly nonlinear dHLS structure with the $\rho$ and $\omega$ mesons decoupling from the nucleons. This transmutation is highly nonlinear involving singularities. A striking prediction of this procedure is that the vector-meson--nucleon vector coupling goes to zero at the dilaton limit. This simply means that as the dilaton limit is approached as density increases, two hitherto unexpected phenomena could occur. Firstly the well-known $\omega$-nucleon interaction known to be repulsive at low density should get strongly suppressed at high density. Secondly the nuclear symmetry energy denoted in the literature as $E_{sym}$ that encodes the energy cost in the excess of neutrons in compact-star systems should also get weaker. An immediate consequence would be that the EoS of dense matter, particularly of compact-star matter, will be softened at higher density. This result is a distinctive feature of the HLS structure of the model that is not present in the absence of HLS vector mesons. An intriguing question is what effect this ``quenching" of the repulsive core will have on the recently observed 1.97 $M_\odot$ neutron star. This question is highly pertinent to the recent description of the 1.97 $M_\odot$ neutron star in terms of a three-layered structure of the compact star consisting of nuclear matter, kaon condensed nuclear matter and strange-quark matter~\cite{KLR}. The suppression of repulsion or effectively an attraction at high density will clearly have an important impact on stabilizing 2-solar-mass stars. This issue will be addressed elsewhere.

\section{The HLS model with dilatons}
\label{sec:hls}

Following the two-component concept for dilatons proposed in~\cite{miransky},
the dilaton potential written in terms of
soft $\chi_s$ and hard $\chi_h$ components
\begin{equation}
V(\chi) = V_s(\chi_s) + V_h(\chi_h)\,,
\end{equation}
will be assumed to have a negligible mixing between soft and hard sectors
in order to avoid an undesirably strong coupling of the glueball to pions.
The expectation value of $\chi_s$ is assumed to vanish when chiral symmetry is
restored~\cite{LeeRho}, whereas the one of $\chi_h$ remains finite, representing the ``explicit breaking" of conformal invariance, i.e., the scale anomaly in QCD. It was shown in ~\cite{LeeRho} with an HLS Lagrangian that the soft dilaton
plays an important role in the emergence of a half-skyrmion phase at high
density where a skyrmion turns into two half skyrmions ~\cite{half}.
In the subsequent sections we construct an effective theory for the soft
dilaton\footnote{Unless otherwise stated, we will denote the soft dilaton simply by $\chi$ while the hard component which plays no role in taking the dilaton limit but figures as the source for $m_0$ in the mirror assignment will be kept as $\chi_h$.}, pions and vector mesons to go from non-linear basis (HLS) to a linear basis, which enables us to deal  readily with the scalar degree of freedom near the chiral symmetry restoration.

The 2-flavored HLS Lagrangian\footnote{In this paper, we consider $N_f=2$, with the $N_f=3$ case taken up later when kaon dynamics is addressed.} is based on
a $G_{\rm{global}} \times H_{\rm{local}}$ symmetry,
where $G_{\rm global}=[SU(2)_L \times SU(2)_R]_{\rm global}$
is the chiral symmetry and
$H_{\rm local}=[SU(2)_V]_{\rm local}$
is the HLS.
The whole symmetry $G_{\rm global}\times H_{\rm local}$
is spontaneously broken to a diagonal $SU(2)_V$.
The basic quantities are
the HLS gauge boson, $V_\mu$,
and
two matrix valued variables $\xi_L$, $\xi_R$,
which are combined in a
$2 \times 2$ special-unitary matrix
$U = \xi_L^\dagger \xi_R$.
These variables are parameterized as
\begin{equation}
\xi_{L,R}(x)=e^{i\sigma (x)/{F_\sigma}}e^{\mp i\pi (x)/{F_\pi}}\,,
\end{equation}
where $\pi = \pi^a T_a$ denotes the pseudoscalar
Nambu-Goldstone (NG)
bosons associated with the spontaneous symmetry breaking of
$G_{\rm{global}}$ chiral symmetry,
and $\sigma = \sigma^a T_a$ denotes
the NG bosons associated with
the spontaneous breaking of $H_{\rm{local}}$~\footnote{
This $\sigma$ has nothing to do with the scalar meson
in the linear sigma model, but it is the longitudinal part of
the vector meson.
}.
The $\sigma$ is absorbed into the HLS gauge boson through
the Higgs mechanism and the gauge boson acquires its mass.
$F_\pi$ and $F_\sigma$ are the decay constants
of the associated particles.

The fundamental objects are the Maurer-Cartan 1-forms
defined by
\begin{eqnarray}
\hat{\alpha}_{\perp }^{\mu}
&=& \frac{1}{2i}\left[ D^\mu\xi_R \cdot \xi_R^{\dagger}
{}- D^\mu\xi_L \cdot \xi_L^{\dagger} \right]\,,
\nonumber\\
\hat{\alpha}_{\parallel}^{\mu}
&=& \frac{1}{2i}\left[ D^\mu\xi_R \cdot \xi_R^{\dagger}
{}+ D^\mu\xi_L \cdot \xi_L^{\dagger} \right]\,.
\end{eqnarray}
where
the covariant derivatives of $\xi_{L,R}$ are given by
\begin{eqnarray}
&&
D_\mu \xi_L
 = \partial_\mu\xi_L - iV_\mu\xi_L + i\xi_L{\cal{L}}_\mu\,,
\nonumber\\
&&
D_\mu \xi_R
 = \partial_\mu\xi_R - iV_\mu\xi_R + i\xi_R{\cal{R}}_\mu\,,
\end{eqnarray}
with ${\cal{L}}_\mu$ and ${\cal{R}}_\mu$ being the external
gauge fields introduced by gauging $G_{\rm{global}}$.
The Lagrangian with lowest derivatives is given by~\cite{HLStree}
\begin{equation}
{\mathcal L}_M
= F_\pi^2\mbox{tr}\left[ \hat{\alpha}_{\perp\mu}
  \hat{\alpha}_{\perp}^{\mu} \right]
{}+ F_\sigma^2\mbox{tr}\left[ \hat{\alpha}_{\parallel\mu}
  \hat{\alpha}_{\parallel}^{\mu} \right]
{}- \frac{1}{2g^2}\mbox{tr}\left[ V_{\mu\nu}V^{\mu\nu} \right]\,,
\end{equation}
where $g$ is the HLS gauge coupling
and the field strengths are defined by
$V_{\mu\nu} = \partial_\mu V_\nu - \partial_\nu V_\mu
{}- i\left[ V_\mu, V_\nu \right]\,$.
One finds the vector meson mass as
\begin{equation}
m_V = g\,F_\sigma\,.
\end{equation}
The nucleon part with HLS is given by~\cite{HLStree}
\begin{equation}
{\mathcal L}_N
= \bar{N}(i\Slash{D} - m_N ) N
{}+ g_A \bar{N}\Slash{\hat{\alpha}}_\perp\gamma_5 N
{}+ g_V \bar{N}\Slash{\hat{\alpha}}_\parallel N\,,
\end{equation}
with the covariant derivative $D_\mu = \partial_\mu - iV_\mu$
and dimensionless parameters $g_A$ and $g_V$.

Conformal invariance can be embedded in chiral Lagrangians by
introducing a scalar field $\tilde{\chi}$ via $\chi = F_\chi\tilde{\chi}$
and $\kappa = (F_\pi/F_\chi)^2$~\cite{vanKolck}. The HLS Lagrangian
with a dilaton potential describing the scale anomaly~\cite{schechter}
is extended to be
\begin{align}
&{\mathcal L}
= {\mathcal L}_N + {\mathcal L}_M
{}+ {\mathcal L}_\chi\,,
\\
&{\mathcal L}_N
= \bar{N}i\Slash{D}N
{}- \frac{\sqrt{\kappa}}{F_\pi} m_N \bar{N}N\chi
{}+ g_A \bar{N}\Slash{\hat{\alpha}}_\perp\gamma_5 N
{}+ g_V \bar{N}\Slash{\hat{\alpha}}_\parallel N\,,
\\
&{\mathcal L}_M
= \kappa \chi^2\, \mbox{tr}
\left[ \hat{\alpha}_{\perp\mu}\hat{\alpha}_\perp^\mu \right]
{}+ a \kappa \chi^2\, \mbox{tr}
\left[ \hat{\alpha}_{\parallel\mu}\hat{\alpha}_\parallel^\mu \right]
{}- \frac{1}{2g^2}\mbox{tr}\left[ V_{\mu\nu} V^{\mu\nu} \right]\,,
\\
&{\mathcal L}_\chi
= \frac{1}{2}\partial_\mu\chi \cdot \partial^\mu\chi
{}+ \frac{\kappa m_\chi^2}{8 F_\pi^2}
\left[ \frac{1}{2}\chi^4
{}- \chi^4\ln\left( \frac{\kappa\chi^2}{F_\pi^2}\right)\right]\,,
\end{align}
where $a = (F_\sigma/F_\pi)^2$ and $m_\chi$ is the mass of the dilaton.

\section{Linearization of the model}
\label{sec:linear}

Near chiral symmetry restoration the quarkonium component of the
dilaton field becomes a scalar mode which forms with pions an O(4)
quartet~\cite{vanKolck}. This can be formulated by making a transformation
of a non-linear chiral Lagrangian to a linear basis exploiting the dilaton limit. One can think of going to the dilaton limit as going toward the chiral restoration point. It should however be stressed that how the process takes place in going to that limit cannot be addressed. We will simply take the effective Lagrangian that results in the dilaton limit as the Lagrangian relevant in the vicinity of chiral restoration.
In this section we derive a linearized Lagrangian assuming two different
chirality assignments to the positive and negative nucleons. Models with
the ``naive" assignment
describe the nucleon mass which is entirely generated by spontaneous chiral symmetry breaking, whereas the mirror assignment
allows an explicit mass term consistently with chiral
invariance~\cite{dk,mirror}.

\subsection{The ``naive" model}
\label{ssec:naive}

Following \cite{vanKolck} we introduce new fields as
\begin{eqnarray}
\Sigma
&=& U\chi\sqrt{\kappa} = \xi_L^\dagger\xi_R \chi\sqrt{\kappa}
= s + i\vec{\tau}\cdot\vec{\pi}\,,
\\
{\mathcal N}
&=& \frac{1}{2}\left[ \left( \xi_R^\dagger + \xi_L^\dagger \right)
{}+ \gamma_5\left( \xi_R^\dagger - \xi_L^\dagger \right) \right] N\,,
\end{eqnarray}
with the Pauli matrices $\vec{\tau}$ in the isospin space.
The linearized Lagrangian includes terms which generate singularities,
negative powers of $\mbox{tr}\left[ \Sigma\Sigma^\dagger\right]$,
in chiral symmetric phase. Those terms carry the following factor:
\begin{equation}
X_N = g_V - g_A\,,
\quad
X_\chi = 1-\kappa\,.
\end{equation}
Assuming that nature disallows any singularities in the case considered, we require that they be absent in the Lagrangian, i.e.
$X_N = X_\chi = 0$. We find $\kappa = 1$ and $g_A = g_V$.
A particular value, $g_V = g_A = 1$, recovers the large $N_c$
algebraic sum rules~\cite{vanKolck}. Thus, we adopt the dilaton
limit as
\begin{equation}
\kappa = g_A = g_V = 1\,.
\end{equation}

\begin{widetext}
In this limit one finds
\begin{eqnarray}
{\mathcal L}
&=& \bar{\mathcal N}i\Slash{\partial}{\mathcal N}
{}- \frac{m_N}{2F_\pi}\bar{\mathcal N}
\left[ \Sigma + \Sigma^\dagger
{}+ \gamma_5\left( \Sigma - \Sigma^\dagger\right)
\right]{\mathcal N}
{}+ \frac{1}{4}\mbox{tr}
\left[ \partial_\mu\Sigma \cdot \partial^\mu\Sigma^\dagger\right]
\nonumber\\
&&
{}+ \frac{a}{2i}\mbox{tr}\left[
\left( \Sigma\partial_\mu\Sigma^\dagger + \Sigma^\dagger\partial_\mu\Sigma
\right) V^\mu \right]
{}+ \frac{a}{2}\mbox{tr}\left[ \Sigma\Sigma^\dagger \right]
\mbox{tr}\left[ V_\mu V^\mu \right]
{}- \frac{1}{2g^2}\mbox{tr}\left[ V_{\mu\nu}V^{\mu\nu}\right]
\nonumber\\
&&
{}+ \frac{m_s^2}{64 F_\pi^2}
\left( \mbox{tr}\left[ \Sigma\Sigma^\dagger\right]\right)^2
{}- \frac{m_s^2}{32 F_\pi^2}
\left( \mbox{tr}\left[ \Sigma\Sigma^\dagger\right]\right)^2
\ln\left( \frac{\mbox{tr}\left[ \Sigma\Sigma^\dagger\right]}{2F_\pi^2}
\right)\,,
\label{naivelag}
\end{eqnarray}
where the unitary gauge $\sigma = 0$ is taken and the dilaton mass
$m_\chi$ is replaced with the mass of the effective scalar mode $m_s$.
\end{widetext}

A noteworthy feature of the dilaton-limit Lagrangian (\ref{naivelag}) is that the vector mesons decouple from the nucleons while their coupling to the Goldstone bosons remains. As announced in Introduction, this has two striking new predictions. Taking the dilaton limit drives the Yukawa interaction to vanish as $g_{VN}^2=(g\,(1-g_V))^2\rightarrow 0$ for $V=\rho, \omega$ for any finite value of $g$. In HLS for the meson sector, the model has the vector manifestation (VM) fixed point as one approaches chiral restoration, so the HLS coupling $g$ also tends to zero proportional to the quark condensate. It thus follows that combined with the VM, the coupling $g_{V N}$ will tend to vanish rapidly near the phase transition point. In nuclear forces, what is effective is the ratio $g_{VN}^2/m_V^2$ which goes as $(1-g_V)^2$. This means that (1) the two-body repulsion which holds two nucleons apart at short distance will be suppressed in dense medium and (2) the symmetry energy going as $S_{sym}\propto g_{\rho N}^2$ will also get suppressed. As a principal consequence, the EoS at some high density approaching the dilaton limit will become softer {\it even without such exotic happenings as kaon condensation or strange quark matter}.

\begin{figure}
\begin{center}
\includegraphics[width = 5cm]{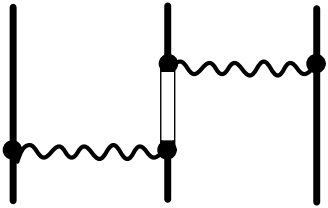}
\end{center}
\caption{
Three-body interaction with the omega-meson exchange. When the intermediate state in the middle nucleon leg is higher-lying than the nucleon, it can become an irreducible 3-body force in the sense defined in chiral perturbation theory in a form of contact interaction when the $\omega$ fields are integrated out.}
\label{3Nomega}
\end{figure}

It has been argued that the short-range point-like three-neutron force essentially
constrains the maximum mass of the neutron star~\cite{3N}. In the ``naive" assignment, we can have such a three-body force from the $\omega$-exchange graph given in Fig.~\ref{3Nomega}. It was suggested in ~\cite{holt} that the same three-body force is predominantly responsible for the suppressed Gamow-Teller matrix element accounting for the long life-time of $^{14}$C. Since the same suppression can be explained very well by Brown-Rho scaling in the nuclear tensor forces without three-body forces~\cite{holtetal}, there may be considerable overlap between the various mechanisms evoked for the process.\footnote{In contrast to \cite{holt} where the contact interaction is seen to play a key role in the GT suppression, a calculation using {\it ab initio} no-core shell model (NCSM)~\cite{vary} finds that the requisite suppression is primarily driven by the long-range two-pion exchange three-body force and not by the contact interaction. The variety of different mechanisms that are seemingly successful are clearly not independent of each other in nuclear structure physics, pointing to the subtlety in which chiral symmetry can be manifested in nuclear medium~\cite{MR-GEBfest}.} Similar overlap may be at work for the EoS of the neutron stars and it may be dangerous to draw conclusions based on one particular model or scheme. In the present scheme with the dilaton encoding the scaling property, both three-body forces and scaling properties can be -- and should be -- consistently taken into account.

In the present scheme, the shortest-range component of the three-body forces is given by the graph of Fig.~\ref{3Nomega}. The intermediate states entering in the middle nucleon line should be higher-lying than nucleon and hence could be integrated out. The resulting effective $NN\omega\omega$ vertex is expected to be smooth-varying in density, remaining finite in the dilaton limit. Therefore the three-body potential of Fig.~\ref{3Nomega} will carry the factor $g_{\omega N}^2$ that vanishes in the dilaton limit. The one-pion exchange three-body force involving a contact two-body force will also get suppressed as $\sim g_{\omega N}^2$.
Thus only the longest-range two-pion exchange three-body forces will remain operative at large density in compact stars. How this intricate mechanism affects the EoS at high density is a challenge issue to resolve.

\subsection{The mirror model}
\label{ssec:mirror}

The Lagrangian of mirror nucleons in the non-linear realization without
vector mesons was considered in~\cite{nemoto}. Its HLS-extended form
is found to be
\begin{align}
{\mathcal L}_N
=
&\bar{B}i\Slash{D}B
{}+ g_A\bar{B}\rho_3\Slash{\hat{\alpha}}_\perp\gamma_5 B
{}+ g_V\bar{B}\Slash{\hat{\alpha}}_\parallel B
\nonumber\\
&
{}- g_1 F_\pi \bar{B}B + g_2 F_\pi \bar{B}\rho_3 B
{}- im_0 \bar{B}\rho_2\gamma_5 B\,,
\end{align}
where $B = (B_1, B_2)^T$ denotes the nucleon doublet in the chiral
eigenstate, the $\rho_i$ are the Pauli matrices in the parity pair space,
and the mass parameters $g_{1,2}$ and $m_0$.
Implementing the dilaton field, one obtains
\begin{align}
{\mathcal L}_N
=
&\bar{B}i\Slash{D}B
{}+ g_A\bar{B}\rho_3\Slash{\hat{\alpha}}_\perp\gamma_5 B
{}+ g_V\bar{B}\Slash{\hat{\alpha}}_\parallel B
\nonumber\\
&
{}- g_1 \sqrt{\kappa}\chi \bar{B}B
{}+ g_2 \sqrt{\kappa}\chi \bar{B}\rho_3 B
{}- im_0 \frac{\chi_h}{F_{\chi_h}} \bar{B}\rho_2\gamma_5 B\,,
\end{align}
where we require that the (broken) scale symmetry is possessed by the hard
dilaton in the last term since the nucleon mass becomes $m_0$ at
chiral symmetry restoration and can be traced back to the non-vanishing
gluon condensate in symmetric phase. In this way we are attributing the origin of $m_0$ to the hard component of the gluon condensate, which is chiral invariant. Of course the origin of $m_0$ could be something else but at this moment, we have no idea as to what that could be.

We linearize the Lagrangian in terms of $\Sigma = U\chi\sqrt{\kappa}$
and the new nucleon fields introduced by
\begin{equation}
\psi_{1,2}
= \frac{1}{2}\left[ \left( \xi_R^\dagger + \xi_L^\dagger \right)
{}\pm \gamma_5\left( \xi_R^\dagger - \xi_L^\dagger \right) \right] B_{1,2}\,.
\end{equation}
As in the ``naive" model, singularities are present in the terms carrying
the same factor:
\begin{equation}
X_N = g_V - g_A\,,
\quad
X_\chi = 1 - \kappa\,.
\end{equation}
The dilaton limit is therefore unchanged by the mirror baryons
and, adapting to the sum rules in large $N_c$, one arrives at
\begin{equation}
\kappa = 1\,,
\quad
g_V = g_A = 1\,.
\end{equation}
This leads to the linearized baryonic Lagrangian as
\begin{align}
{\mathcal L}_N
&=
\bar{\psi}i\Slash{\partial}\psi
{}- \frac{g_1}{2}\bar{\psi}\left[ \left( \Sigma + \Sigma^\dagger \right)
{}+ \rho_3\gamma_5 \left( \Sigma - \Sigma^\dagger\right)\right]\psi
\nonumber\\
&
{}+ \frac{g_2}{2}\bar{\psi}\left[ \rho_3\left( \Sigma + \Sigma^\dagger
\right)
{}+ \gamma_5 \left( \Sigma - \Sigma^\dagger\right)\right]\psi
{}- i\bar{m}_0 \bar{\psi}\rho_2\gamma_5\psi\,,
\end{align}
with $\bar{m}_0 = (\chi_h/F_{\chi_h})m_0$.
The mass term is diagonalized by the mass eigenstates
of the parity doubled nucleons, ${\cal N}_+$ and ${\cal N}_-$ via
\begin{equation}
\begin{pmatrix}
{\cal N}_+
\\
{\cal N}_-
\end{pmatrix}
=
\frac{1}{\sqrt{2\cosh\delta}}
\begin{pmatrix}
e^{\delta/2} & \gamma_5 e^{-\delta/2}
\\
\gamma_5 e^{-\delta/2} & -e^{\delta/2}
\end{pmatrix}
\begin{pmatrix}
\psi_1
\\
\psi_2
\end{pmatrix}\,,
\end{equation}
with $\sinh\delta = g_1 s /\bar{m}_0$.
The Lagrangian in this basis is thus given by
\begin{align}
{\mathcal L}_N
=
&\bar{\cal N}i\Slash{\partial}{\cal N} - \bar{\cal N}\hat{M}{\cal N}
{}- g_1\bar{\mathcal N}\left(
\hat{G}\tilde{s} + \rho_3\gamma_5 i\tau\cdot\vec{\tilde{\pi}}
\right) {\mathcal N}
\nonumber\\
&
{}+ g_2\bar{\mathcal N}\left(
\rho_3 \tilde{s} + \hat{G}\gamma_5 i\tau\cdot\vec{\tilde{\pi}}
\right) {\mathcal N}\,,
\label{mirrorlag}
\end{align}
where $\tilde{s}$ and $\tilde{\pi}$ are fluctuations around their
expectation values, the matrix $\hat{G}$ is defined by
\begin{equation}
\hat{G} =
\begin{pmatrix}
\tanh\delta & \gamma_5/\cosh\delta \\
-\gamma_5/\cosh\delta & \tanh\delta
\end{pmatrix}\,,
\end{equation}
and the mass matrix $\hat{M} = {\rm{diag}}(m_+,m_-)$ with
\begin{equation}
m_\pm = \mp g_2 \langle s \rangle
{}+ \sqrt{(g_1 \langle s \rangle)^2 + \bar{m}_0^2}\,.
\end{equation}
Note that the new axial-coupling of the nucleon,
\begin{equation}
\bar{g}_A = \tanh\delta\,,
\end{equation}
is obtained and the corresponding Goldberger-Treiman relation
is satisfied: $g_{\pi N_+ N_+} = \bar{g}_A m_+/\langle s \rangle$.

The softening of EoS at large density in the mirror model is quite similar to the case of the ``naive" model. The suppression of the vector coupling is of the same form:
\begin{equation}
g_{V N} = g\,(1-g_V) \to 0\,.
\end{equation}
That the quenching of the short-range repulsion is independent of the chirality assignment of the nucleon is indicative of a universality of the short-distance interaction. Now the consequence on this coupling will be sensitive to how chiral symmetry is restored in the given scenario. In the ``naive" HLS theory, the chiral symmetry is restored as the VM
characterized by the fixed point of
renormalization group equations, $g \to 0$ and $a \to 1$, which leads to
the massless vector meson as the chiral partner of the pion~\cite{HLSloop},
and remains so in the presence of constituent quarks (or fermions in the ``naive"
assignment).
It is not obvious in the mirror model that in the region where the linearized model describes
the melting chiral condensate, $\langle s \rangle \to 0$, the dropping
HLS gauge coupling remains the fixed point. Nevertheless, one would expect in the mirror case as well
a reduced $g$ before reaching the dilaton limit as a tendency
of the VM. Therefore, the $g_{V N}$ could also be thought to be weakened toward
the restoration point, leading to a softer EoS of dense matter in an analogous way.

It seems natural to expect that the source for non-zero $m_0$ is in the hard dilaton condensate so far ignored in dealing with the part of the nucleon mass dynamically generated. How large is $m_0$ at the chiral symmetry restoration? Here we make a rough estimate from thermodynamic considerations.

Assuming a second-order chiral phase transition, i.e.
$\langle s \rangle \sim 0$, thermodynamics around the critical point
is described by the following potential under the mean-field
approximation:
\begin{align}
\Omega
&= 8\int\frac{d^3p}{(2\pi)^3}
T \left[ \ln\left( 1 - n_0 \right)
{}+ \ln\left( 1 - \bar{n}_0 \right)\right]
\nonumber\\
&
{}+ V(\chi_h)
{}- \int\frac{d^3p}{(2\pi)^3}
T \ln\left( 1 + n_h \right)\,,
\end{align}
where $V$ is the potential for the hard dilaton given by
\begin{equation}
V = \frac{1}{4}B_h\left( \frac{\chi_h}{F_{\chi_h}}\right)^4
\left[ \ln\left(\frac{\chi_h}{F_{\chi_h}}\right)^4 - 1\right]\,,
\end{equation}
and the statistical distribution functions are
\begin{align}
n_0 &= \frac{1}{e^{(E_0 - \mu)/T} + 1}\,,
\nonumber\\
\bar{n}_0 &= \frac{1}{e^{(E_0 + \mu)/T} + 1}\,,
\nonumber\\
n_h &= \frac{1}{e^{E_h/T} - 1}\,.
\end{align}
The energy $E = \sqrt{\vec{p}^2 + m^2}$ of the corresponding particles
is given for the parity doubled nucleons with
$m_\pm = \bar{m}_0 = m_0\left(\chi_h/F_{\chi_h} \right)$, and
for the hard dilaton with its mass introduced by
\begin{equation}
m_{\chi_h}^2 = \frac{\partial^2 V}{\partial\chi_h^2}
= B_h\left(\frac{\chi_h}{F_{\chi_h}}\right)^2\frac{1}{F_{\chi_h}^2}
\left[ 3\ln\left(\frac{\chi_h}{F_{\chi_h}}\right)^4 + 4\right]\,.
\label{hardmass}
\end{equation}

In what follows, we restrict our analysis to a hot system at zero chemical
potential where gluodynamics is well guided by lattice QCD.
One obtains the gap equation for a nontrivial $\chi_h$ from the stationary
condition, $\frac{\partial\Omega}{\partial\chi_h}=0$, as
\begin{align}
&
16\int\frac{d^3p}{(2\pi)^3}\frac{m_0^2}{E_0}n_0
{}+ B_h\left(\frac{\chi_h}{F_{\chi_h}}\right)^2
\ln\left(\frac{\chi_h}{F_{\chi_h}}\right)^4
\nonumber\\
&
{}+ \int\frac{d^3p}{(2\pi)^3}\frac{B_h}{E_h F_{\chi_h}^2}
\left[ 3\ln\left(\frac{\chi_h}{F_{\chi_h}}\right)^4 + 10\right] n_h
= 0\,.
\label{gapeqhard}
\end{align}
The gluon condensate calculated on a lattice in the presence of
dynamical quarks is known to be~\cite{miller}
\begin{equation}
\langle G_{\mu\nu}G^{\mu\nu}\rangle_{T_{\rm ch}}
\simeq \frac{1}{2}\langle G_{\mu\nu}G^{\mu\nu}\rangle_{T=0}\,,
\end{equation}
at pseudo-critical temperature $T_{\rm ch} \sim 170$ MeV.
We thus adopt the bag constant and mass for the hard dilaton as
\begin{equation}
B_h(T_{\rm ch}) = \frac{1}{2}B(T=0)\,,
\quad
m_{\chi_h}^2 = \frac{1}{2}m_G^2\,,
\end{equation}
using the bag constant $B$ and the glueball mass $m_G$ in vacuum.
With the empirical numbers for those parameters,
$\langle G_{\mu\nu}G^{\mu\nu}\rangle_{T=0} = 0.0012$ GeV$^4$~\cite{shifman},
$B=(0.4\,\mbox{GeV})^4$~\cite{narison} and $m_G = 1.7$ GeV~\cite{sexton},
Eqs.~(\ref{hardmass}) and (\ref{gapeqhard}) determine $m_0$ for a given
$F_{\chi_h}$. In QCD trace anomaly exists at higher temperature and thus
the expectation value of $\chi_h$ is supposed to be fairly close to
$F_{\chi_h}$. To make a rough estimate, we take $\langle \chi_h \rangle
= 0.99\,F_{\chi_h}$. This gives $m_0 = 210$ MeV as a solution favored
thermodynamically. This is in agreement with the estimate made in vacuum phenomenology~\cite{PLRS}. The nucleon in the mirror model stays massive
at chiral symmetry restoration, so a different EoS from that in the ``naive''
model would be expected. This issue and more realistic estimate of $m_0$
will be reported in a subsequent publication.

\section{Mixing between quarkonium and tetraquarks}
\label{sec:mixing}

In taking the dilaton limit, we went from a low-density state with the dilaton, a chiral singlet, to a high-density state with the $\sigma$, the fourth component of the chiral four vector. The former is appropriate for low-energy nuclear physics resembling Walecka mean field model and the latter is for chiral phase restoration. How this change-over takes place is not explained. How can this happen in the language of QCD?

In general, in the scalar sector of low-mass hadrons, we expect to have scalar quarkonium, tetraquark states and glueballs. They will naturally be all mixed. It is reasonable to assume that the mixing between soft
and hard gluon sectors is negligible as is done in the dilaton potential.
The soft dilaton $\chi_s$ is invariant under the $U_A(1)$
transformation, while the 2-quark and 4-quark states are not.
The entire dilaton, $\chi_s + \chi_h$, is chiral-singlet. Since we
are assuming no mixing between the soft and hard dilatons, what we should consider
is the mixing among the 2-quark, 4-quark states and $\chi_s = \chi$.
Once we make a linearization with $\Sigma = U\chi\sqrt{\kappa}$, the scalar
mode appearing in the Lagraingian is a mixture of
the quarkomium and soft dilaton, and we cannot make a separation of them. For simplicity,
we will simply ignore this subtlety, and consider the mixing between the quarkonium $s$
and the tetraquark fields $\psi$, thus restricting to a two-level system.

\begin{figure}
\begin{center}
\includegraphics[width = 8cm]{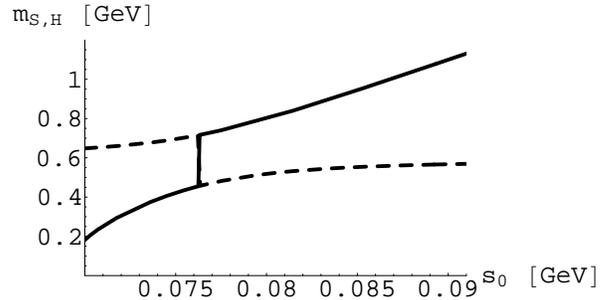}
\end{center}
\caption{
A sketch of the behavior of $m_S$ (solid) and $m_H$ (dotted)
as functions of the chiral condensate $s_0$. For an illustrative
purpose, the parameters are set to be $h=1$ GeV, $m_s = 1.2$ GeV,
$m_\psi = 0.6$ GeV, $F_\pi = 93$ MeV.
}
\label{scalarmass}
\end{figure}

The relevant mesonic potential is~\cite{4Qschechter,4QT}
\begin{align}
{\mathcal U}
&= \frac{1}{2}m_\psi^2 \psi^2 - h\psi\left( s^2 + \vec{\pi}^2 \right)
\nonumber\\
&
{}- \frac{m_s^2}{16 F_\pi^2}\left( s^2 + \vec{\pi}^2 \right)^2
\left[ 1 - 2\ln\left( \frac{s^2 + \vec{\pi}^2}{F_\pi^2} \right) \right]\,,
\end{align}
with $h$ being the mixing strength of $s$ and $\psi$ fields.
Shifting the fields around their expectation values, $s_0$ and $\psi_0$,
the potential reads
\begin{equation}
{\mathcal U}
= \frac{1}{2}\bar{m}_s^2 s^2 + \frac{1}{2}m_\psi^2 \psi^2
{}- 2hs_0 s\psi + \cdots\,,
\end{equation}
where ellipses stand for the terms including the higher fields than cubic, and
\begin{equation}
\bar{m}_s^2
= m_s^2\frac{s_0^2}{F_\pi^2}\left[ 1
{}+ 3\ln\left( \frac{s_0}{F_\pi}\right) \right]
{}- 2h\psi_0\,.
\end{equation}
The quadratic term thus becomes
\begin{equation}
{\mathcal U}^{(2)}
= \frac{1}{2}\left( s\,, \psi \right)
\begin{pmatrix}
\bar{m}_s^2 & -2hs_0 \\
-2hs_0 & m_\psi^2
\end{pmatrix}
\begin{pmatrix}
s \\
\psi
\end{pmatrix}\,.
\end{equation}
The mass eigenstates are introduced with a rotation matrix as
\begin{eqnarray}
&&
\left(
\begin{matrix}
S
\\
H
\end{matrix}
\right)
=
\left(
\begin{matrix}
\cos\theta & \sin\theta
\\
-\sin\theta & \cos\theta
\end{matrix}
\right)
\left(
\begin{matrix}
s
\\
\psi
\end{matrix}
\right)\,,
\end{eqnarray}
with the angle
\begin{eqnarray}
\tan\left( 2\theta \right)
= \frac{4hs_0}
{m_\psi^2 - \bar{m}_s^2}\,.
\end{eqnarray}
The masses of scalar mesons are give by
\begin{eqnarray}
m_S^2 = \bar{m}_s^2\cos^2\theta + m_\psi^2\sin^2\theta
{}- 2hs_0\sin(2\theta)\,,
\nonumber\\
m_H^2 = m_\psi^2\cos^2\theta + \bar{m}_s^2\sin^2\theta
{}+ 2hs_0\sin(2\theta)\,.
\end{eqnarray}
Fig.~\ref{scalarmass} shows a schematic structure of the masses versus the chiral condensate.
One observes a level crossing between the two scalar states when
$\theta = \pi/4$. The two-quark component of $m_S$ gets more
dominant for smaller $s_0$ and eventually the $S$ state becomes
massless at chiral symmetry restoration whereas $m_H$ is dominated
by the four-quark state and stays massive.

We should point out several caveats in the reasoning given above.

This consideration can be only qualitative since due to
the specific form of the dilaton potential, the models gives a first
order transition. In fact, $m_S (s_0)$ is non-monotonic and becomes
unphysical below $s_0 = 68.9$ MeV within the above setup.
What we must have is a second-order with scalars at finite $T$
and/or $\mu$. Therefore, the current dilaton potential needs to be
modified in the presence of matter where a temperature and a chemical
potential are additional scales responsible for the trace anomaly,
other than $\Lambda_{\rm QCD}$.

The mixing strength $h$ may be determined in matter-free space with the
known spectroscopy for the scalars. This requires us to extend the
model to three flavors, which is beyond the scope of the paper and will
be reported elsewhere.

\section{Role of axial-vector mesons}
\label{sec:a1}

Up to this point we have ignored the axial-vector mesons that figure in
the mended symmetry.
As long as their masses are greater than the masses of other mesons, the axial-vector mesons can be integrated out. However the mended
symmetry of Weinberg implies that all mesons within the given multiplet become degenerate, and massless when the chiral
symmetry is restored. At that point one must deal with the axial-vector mesons on the same footing with the others. In this section, we give a brief discussion of how the axial vector mesons can be incorporated into the HLS framework. Incorporating baryons, both in the ``naive" assignment and in the mirror assignment, is straightforward and hence will not be explicited.

Axial-vector mesons are introduced by generalizing $H_{\rm local}$
to $G_{\rm local}$ (GHLS) so that the entire symmetry of the
theory becomes $G_{\rm global} \times G_{\rm local}$~\cite{HLStree,GHLS}.
The Maurer-Cartan 1-forms are defined by
\begin{eqnarray}
\hat{\alpha}_{L,R}^\mu = D^\mu\xi_{L,R}\cdot\xi_{L,R}^\dagger /i\,,
\quad
\hat{\alpha}_M^\mu = D^\mu\xi_M\cdot\xi_M^\dagger /(2i)\,,
\end{eqnarray}
where $U = \xi_L^\dagger \xi_M \xi_R$ and
the covariant derivatives of $\xi_{L,R,M}$ are given by
\begin{eqnarray}
&&
D_\mu \xi_L
 = \partial_\mu\xi_L - iL_\mu\xi_L\,,
\nonumber\\
&&
D_\mu \xi_R
 = \partial_\mu\xi_R - iR_\mu\xi_R\,,
\nonumber\\
&&
D_\mu \xi_M
 = \partial_\mu\xi_M - iL_\mu\xi_M + i\xi_M R_\mu\,,
\end{eqnarray}
with the GHLS gauge bosons, $L_\mu$ and $R_\mu$,
identified with the vector and axial-vector mesons as
$V_\mu = (R_\mu + L_\mu)/2$ and
$A_\mu = (R_\mu - L_\mu)/2$.
Imposing the Weinberg sum rules~\footnote{
 This corresponds to the theory space locality~\cite{locality}, i.e. the mixing of
 left and right chirality is generated only through gauge bosons.
}, the Lagrangian of the meson sector is given by~\cite{GHLSloop}
\begin{align}
{\mathcal L}_M
&=
aF^2 \left( \mbox{tr}\left[ \hat{\alpha}_{\perp\mu}
  \hat{\alpha}_{\perp}^{\mu} \right]
{}+ \mbox{tr}\left[ \hat{\alpha}_{\parallel\mu}
  \hat{\alpha}_{\parallel}^{\mu} \right]
\right)
{}+ cF^2 \mbox{tr}\left[ \hat{\alpha}_{M\mu}
  \hat{\alpha}_{M}^{\mu} \right]
\nonumber\\
&
{}- \frac{1}{2g^2}\mbox{tr}\left[ V_{\mu\nu}V^{\mu\nu} \right]
{}- \frac{1}{2g^2}\mbox{tr}\left[ A_{\mu\nu}A^{\mu\nu} \right]\,,
\end{align}
with a dimension-1 parameter $F$, two dimensionless ones $a$ and $c$
and
$\hat{\alpha}_{\parallel,\perp}^\mu
 = \bigl( \xi_M\hat{\alpha}_R^\mu\xi_M^\dagger
  {}\pm \hat{\alpha}_L^\mu \bigr)/2\,$.
No new ingredients are introduced in coupling to nucleons, so we will focus on mesons only.

Fields for three types of Nambu-Goldstone (NG) bosons,
$\phi_\sigma, \phi_\perp$ and $\phi_p$, are introduced as
\begin{equation}
\xi_{L,R} = e^{i(\phi_\sigma \mp \phi_\perp)}\,,
\quad
\xi_M = e^{2i\phi_p}\,.
\end{equation}
Solving the $\pi$-$A$ mixing the pion field $\phi_\pi$ is found to be
the combination
\begin{equation}
\phi_\pi = \phi_\perp + \phi_p\,,
\end{equation}
while two remaining would-be NG bosons,
$\phi_\sigma$ and
\begin{equation}
\phi_q = \frac{1}{a+c}\left( c\phi_p - a\phi_\perp \right)\,,
\end{equation}
representing the longitudinal vector and axial-vector
degrees of freedom,
are absorbed into the $\rho$ and $a_1$.
The pion decay constant is given by
\begin{equation}
F_\pi^2 = \frac{ac}{a+c} F^2\,.
\end{equation}

Following the same procedure carried out in Section~\ref{sec:linear},
the non-linear GHLS Lagrangian with introducing a soft dilaton is
transformed to its linearized form. Taking the unitary gauge one obtains
\begin{align}
&
{\mathcal L}_M + {\mathcal L}_{\chi\rm{kin}}
=
\frac{1}{4}\mbox{tr}
\left[ \partial_\mu\Sigma \cdot \partial^\mu\Sigma^\dagger\right]
\nonumber\\
&
{}+ \frac{(a+c)^3}{2ac}\mbox{tr}\left[ \Sigma\Sigma^\dagger \right]
\mbox{tr}\left[ A_\mu A^\mu \right]
{}+ \frac{a(a+c)}{2c}\mbox{tr}\left[ \Sigma\Sigma^\dagger \right]
\mbox{tr}\left[ V_\mu V^\mu \right]
\nonumber\\
&
{}+ \frac{a(a+c)}{2ic}\mbox{tr}\left[
\left( \Sigma\partial_\mu\Sigma^\dagger + \Sigma^\dagger\partial_\mu\Sigma
\right) V^\mu \right]
\nonumber\\
&
{}- \frac{1}{2g^2}\mbox{tr}\left[ V_{\mu\nu}V^{\mu\nu}\right]
{}- \frac{1}{2g^2}\mbox{tr}\left[ A_{\mu\nu}A^{\mu\nu}\right]\,.
\end{align}
The vector meson masses in the mean field approximation read
\begin{equation}
m_\rho^2 = \frac{a(a+c)}{c}g^2\langle s \rangle^2\,,
\quad
m_{a_1}^2 = \frac{(a+c)^3}{ac}g^2\langle s \rangle^2\,.
\end{equation}
When chiral symmetry restoration takes place, the $\rho$ and $a_1$
mesons become massless as the chiral condensate is melting,
$\langle s \rangle \to 0$.

\section{Conclusions}
\label{sec:conc}

The basic premise in our line of thinking was that local field degrees of freedom make sense -- and hence the notion of mended symmetry is applicable -- up to the point where the density-driven chiral phase transition takes place. This would preclude strongly first-order transitions or the total breakdown of description in terms of quasiparticles -- such as ``hadron melting" -- in the vicinity of the transition. If it were so, our discussion would be of no meaning. At present, there are no indications that enable us to make a firm statement on that.

Taking the dilaton limit \`a la Beane and van Kolck on a dilaton-implemented hidden local symmetry Lagrangian that we identify with going to near chiral restoration density, we uncover a number of surprising results in both ``naive" and mirror models due to that vector mesons decouple from baryons. One  important prediction is that the repulsion at short distance in nuclear interactions should get suppressed at a density in the vicinity of the dilaton limit. Another hitherto unsuspected result is that the symmetry energy which plays a crucial role in the structure of compact stars also should get suppressed. Put together, they will soften the EoS of compact-star matter at some high density. An interesting possibility is that our mechanism could accommodate an exotica-free nucleon-only EoS (such as AP4 in Fig. 3 of \cite{ozel}) with a requisite softening at higher density that could be compatible with the $1.97 \pm 0.04\,M_\odot$ neutron star data~\cite{2solarmass}.

Now what is known about the mysterious repulsive core?

It is well-established in matter-free space that there is a strong repulsion between two nucleons.  In fact, it is confirmed in lattice gauge calculations~\cite{lattice}. And there are evidences from NN scattering. However the mechanism of the two-body repulsion is mysterious and remains unexplained. It could be a Pauli-exclusion principle effect at the quark level or topological effect in terms of the baryon-number-2 soliton etc.  In effective field theory, it can be explained in terms of an $\omega$ exchange. In fact a similar structure is seen in holographic QCD models where an infinite tower of vector mesons figure~\cite{hcore}. There is no lattice information for three-body forces but model considerations predict similar repulsion for them as well.

When it comes to nuclear matter and denser matter, the situation is totally unclear. What one has learned from nuclear structure studies is that the ``hard core" is not a physical observable in medium, that is, it is not visible. It is shoved under what is known as ``short-range correlation." In fact, nuclear structure approaches anchored on effective field theory and renormalization group show that the ``hard-core" repulsion present in two-nucleon potentials plays no role in low-energy physical observables~\cite{kuo}.

What we find in our model is a surprisingly simple mechanism for taming the hard core in many-body systems. Within the field theory framework we are working with, the short-distance repulsion is suppressed in the background or ``vacuum" defined by density. We cannot say whether and how this mechanism can be related to the ``short-range correlation" of nuclear physics, but it offers a possible way to understand it from the mended symmetry point of view.

Our main observation on the suppressed repulsive interaction is a common
feature in the two different assignments, ``naive" and mirror, of chirality.
The nucleon mass near chiral symmetry restoration exhibits a striking
difference in the two scenarios, and the EoS in the mirror model is supposed
to be stiffer than that in the ``naive" model. How the dilaton-limit suppression of the repulsion -- which seems to be universal independent of the assignments but may manifest itself differently in the two cases -- will affect the EoS for compact stars is an interesting question to investigate.

Finally some comments on the nature of the dialton at low and high densities. We have assumed that taking the dilaton limit effectuates a level crossing between two (or three) levels in such a way that at low density the relevant scalar degree of freedom is a low-mass ($\sim 600$ MeV) chiral singlet effective for binding in nuclei and at high density it is the $\sigma$, the 4-th component of the chiral four vector for $N_f=2$, effective in ``mending" the relevant symmetry. How this can happen has been studied in certain simple models~\cite{4QT} but it is highly likely that the physics involved in such change-over is a lot more intricate. This is evidenced by the indication that a level crossing of a similar nature occurs in scalar channel if one varies the number of colors ($N_c$)~\cite{level-crossing-nc}. A proper understanding will require correlating these and possibly other mechanisms involved in the change-over.

\subsection*{Acknowledgments}

We acknowledge partial support by the WCU project
of the Korean Ministry of Educational Science and Technology
(R33-2008-000-10087-0).
The work of C.S. has been partly supported by the Hessian LOEWE initiative
through the Helmholtz International Center for FAIR (HIC for FAIR).


\end{document}